\documentclass[11pt]{article}
\usepackage{amsmath,amssymb}
\usepackage{array}
\newcount\figureno     \figureno=0
\newdimen\figdim       \figdim=70mm
\def\figureinc{%
   \global\advance\figureno by 1%
}
\def\figcaption#1#2#3{\hbox to #2{\hss{\vbox{\hsize=#2 \parindent=0pt
        {\bf Figure \number\figureno#3 :\ }#1}}\hss}
}

\begin{document}
\baselineskip 100pt
\renewcommand{\baselinestretch}{1.5}
\renewcommand{\arraystretch}{0.666666666}
{\large
\parskip.2in
\numberwithin{equation}{section}
\newcommand{\be}{\begin{equation}}
\newcommand{\ee}{\end{equation}}
\newcommand{\ben}{\begin{equation*}}
\newcommand{\een}{\end{equation*}}
\newcommand{\eqalinb}{\begin{eqnarray}}
\newcommand{\eqaline}{\end{eqnarray}}
\newcommand{\br}{\bar}
\newcommand{\fr}{\frac}
\newcommand{\lm}{\lambda}
\newcommand{\ra}{\rightarrow}
\newcommand{\al}{\alpha}
\newcommand{\bt}{\beta}
\newcommand{\z}{\zeta}
\newcommand{\pa}{\partial}
\newcommand{\hs}{\hspace{5mm}}
\newcommand{\up}{\upsilon}
\newcommand{\bigb}{\hspace{7mm}}
\newcommand{\dg}{\dagger}
\newcommand{\vphi}{\vec{\varphi}}
\newcommand{\ve}{\varepsilon}
\newcommand{\acc}{\\[3mm]}
\newcommand{\dl}{\delta}
\newcommand{\sdil}{\ensuremath{\rlap{\raisebox{.15ex}{$\mskip
6.5mu\scriptstyle+ $}}\subset}}
\newcommand{\sdir}{\ensuremath{\rlap{\raisebox{.15ex}{$\mskip
6.5mu\scriptstyle+ $}}\supset}}
\def\tablecap#1{\vskip 3mm \centerline{#1}\vskip 5mm}
\def\p#1{\partial_#1}
\newcommand{\pd}[2]{\frac{\partial #1}{\partial #2}}
\newcommand{\pdn}[3]{\frac{\partial #1^{#3}}{\partial #2^{#3}}}
\def\DP#1#2{D_{#1}\varphi^{#2}}
\def\dP#1#2{\partial_{#1}\varphi^{#2}}
\def\xh{\hat x}
\newcommand{\Ref}[1]{(\ref{#1})}
\def\ld{\,\ldots\,}

\def\C{{\mathbb C}}
\def\Z{{\mathbb Z}}
\def\R{{\mathbb R}}
\def\mod#1{ \vert #1 \vert }
\def\chapter#1{\hbox{Introduction.}}
\def\Sin{\hbox{sin}}
\def\Cos{\hbox{cos}}
\def\Exp{\hbox{exp}}
\def\Ln{\hbox{ln}}
\def\Tan{\hbox{tan}}
\def\Cot{\hbox{cot}}
\def\Sinh{\hbox{sinh}}
\def\Cosh{\hbox{cosh}}
\def\Tanh{\hbox{tanh}}
\def\Asin{\hbox{asin}}
\def\Acos{\hbox{acos}}
\def\Atan{\hbox{atan}}
\def\Asinh{\hbox{asinh}}
\def\Acosh{\hbox{acosh}}
\def\Atanh{\hbox{atanh}}
\def\frac#1#2{{\textstyle{#1\over #2}}}

\def\ph{\varphi_{m,n}}
\def\phl{\varphi_{m-1,n}}
\def\phr{\varphi_{m+1,n}}
\def\varphil{\varphi_{m-1,n}}
\def\varphir{\varphi_{m+1,n}}
\def\varphit{\varphi_{m,n+1}}
\def\varphib{\varphi_{m,n-1}}
\def\pht{\varphi_{m,n+1}}
\def\phb{\varphi_{m,n-1}}
\def\phbl{\varphi_{m-1,n-1}}
\def\phbr{\varphi_{m+1,n-1}}
\def\phtl{\varphi_{m-1,n+1}}
\def\phtr{\varphi_{m+1,n+1}}
\def\u{u_{m,n}}
\def\ul{u_{m-1,n}}
\def\ur{u_{m+1,n}}
\def\ut{u_{m,n+1}}
\def\ub{u_{m,n-1}}
\def\utr{u_{m+1,n+1}}
\def\ubl{u_{m-1,n-1}}
\def\utl{u_{m-1,n+1}}
\def\ubr{u_{m+1,n-1}}
\def\v{v_{m,n}}
\def\vl{v_{m-1,n}}
\def\vr{v_{m+1,n}}
\def\vt{v_{m,n+1}}
\def\vb{v_{m,n-1}}
\def\vtr{v_{m+1,n+1}}
\def\vbl{v_{m-1,n-1}}
\def\vtl{v_{m-1,n+1}}
\def\vbr{v_{m+1,n-1}}

\def\U{U_{m,n}}
\def\Ul{U_{m-1,n}}
\def\Ur{U_{m+1,n}}
\def\Ut{U_{m,n+1}}
\def\Ub{U_{m,n-1}}
\def\Utr{U_{m+1,n+1}}
\def\Ubl{U_{m-1,n-1}}
\def\Utl{U_{m-1,n+1}}
\def\Ubr{U_{m+1,n-1}}
\def\V{V_{m,n}}
\def\Vl{V_{m-1,n}}
\def\Vr{V_{m+1,n}}
\def\Vt{V_{m,n+1}}
\def\Vb{V_{m,n-1}}
\def\Vtr{V_{m+1,n+1}}
\def\Vbl{V_{m-1,n-1}}
\def\Vtl{V_{m-1,n+1}}
\def\Vbr{V_{m+1,n-1}}
\def\tr{{\rm tr}\,}

\def\a{\alpha}
\def\b{\beta}
\def\g{\gamma}
\def\d{\delta}
\def\ep{\epsilon}
\def\e{\varepsilon}
\def\z{\zeta}
\def\t{\theta}
\def\k{\kappa}
\def\l{\lambda}
\def\s{\sigma}
\def\f{\varphi}
\def\w{\omega}
\def\v{{\hbox{v}}}
\def\u{{\hbox{u}}}
\def\x{{\hbox{x}}}

\newcommand{\ie}{{\it i.e.}}
\newcommand{\cmod}[1]{ \vert #1 \vert ^2 }
\newcommand{\cmodn}[2]{ \vert #1 \vert ^{#2} }
\newcommand{\nhat}{\mbox{\boldmath$\hat n$}}
\nopagebreak[3]
\bigskip

\title{ \bf Invariant solutions of a nonlinear wave equation with a small dissipation obtained via approximate symmetries}
\vskip 1cm

\bigskip
\author{
A.~M. Grundland\thanks{email address: grundlan@crm.umontreal.ca}
\\
Centre de Recherches Math{\'e}matiques, Universit{\'e} de Montr{\'e}al,\\
C. P. 6128, Succ.\ Centre-ville, Montr{\'e}al, (QC) H3C 3J7,
Canada\\ Universit\'{e} du Qu\'{e}bec, Trois-Rivi\`{e}res, CP500 (QC) G9A 5H7, Canada \acc A. J. Hariton\thanks{email
address: hariton@crm.umontreal.ca}
\\
Centre de Recherches Math{\'e}matiques, Universit{\'e} de Montr{\'e}al,\\
C. P. 6128, Succ.\ Centre-ville, Montr{\'e}al, (QC) H3C 3J7,
Canada} \date{}

\maketitle
\begin{abstract}
In this paper, it is shown how a combination of approximate symmetries of a nonlinear wave equation with small dissipations and singularity analysis provides exact analytic solutions. We perform the analysis using the Lie symmetry algebra of this equation and identify the conjugacy classes of the one-dimensional subalgebras of this Lie algebra. We show that the subalgebra classification of the integro-differential form of the nonlinear wave equation is much larger than the one obtained from the original wave equation. A systematic use of the symmetry reduction method allows us to find new invariant solutions of this wave equation.
\end{abstract}





\vspace{3mm}

\newpage

\section{Introduction}

A systematic computational method for constructing an approximate symmetry group for a given system of partial differential equations (PDEs) has been extensively developed by many authors, see e.g. \cite{Ames,Bluman1,Fushchich}. A broad review of recent developments in this subject can be found in such books as G. Bluman and S. Kumei \cite{Bluman2}, P. Olver \cite{Olver}, D. Sattinger and O. Weaver \cite{Sattinger}, B. Rozdestvenskii and N. Janenko \cite{Rozdestvenskii} and V. Baikov, R. Gazizov and N. Ibragimov \cite{Ibragimov,Ibragimov2}. Recently, M. Ruggieri and M. Speciale \cite{Ruggieri} determined the Lie algebras of approximate symmetries of nonlinear wave equations admitting a small perturbative dissipation. They discussed the generators of four different versions of the system of equations associated with the nonlinear wave equation
\begin{equation}
u_{tt}=\left[f(u)u_x\right]_x,
\label{i1}
\end{equation}
where $u(t,x)$ is a function of $t$ and $x$. They considered the following second-order PDE with a small dissipative term:
\begin{equation}
u_{tt}=\left[f(u)u_x\right]_x+\varepsilon\left[\lambda(u)u_t\right]_{xx},
\label{i2}
\end{equation}
where $\varepsilon<<1$ is a small parameter and $f$ and $\lambda$  are smooth functions of $u$. If we suppose that the function $u(t,x)$ can be written as
\begin{equation}
u(t,x,\varepsilon)=u_0(t,x)+\varepsilon u_1(t,x)+{\mathcal O}(\varepsilon^2),
\label{i3}
\end{equation}
where $u_0$ and $u_1$ are smooth functions of $t$ and $x$, then equation (\ref{i2}) becomes the following two equations:
\begin{equation}
u_{0,tt}-f(u_0)u_{0,xx}-f'(u_0)(u_{0,x})^2=0,
\label{bigeq1}
\end{equation}
and 
\begin{equation}
\begin{split}
& u_{1,tt}-f(u_0)u_{1,xx}-f'(u_0)u_{0,xx}u_1-2f'(u_0)u_{0,x}u_{1,x}-f''(u_0)(u_{0,x})^2u_1 \\ &-\lambda''(u_0)(u_{0,x})^2u_{0,t}-\lambda'(u_0)u_{0,xx}u_{0,t}-2\lambda'(u_0)u_{0,x}u_{0,xt}-\lambda(u_0)u_{0,xxt}=0.
\end{split}
\label{bigeq2}
\end{equation}
The Lie symmetry algebra of equations (\ref{bigeq1}) and (\ref{bigeq2}) was identified for three separate cases \cite{Ruggieri}:
\begin{equation}
\begin{split}
(I):& \quad f(u_0)=f_0e^{\frac{1}{p}u_0} , \quad \lambda(u_0)=\lambda_0e^{\frac{1+s}{p}u_0}\\
(II):& \quad f(u_0)=f_0(u_0+q)^{\frac{1}{p}} , \quad \lambda(u_0)=\lambda_0(u_0+q)^{\frac{1+s}{p}-1}\\
(III):& \quad f(u_0)=f_0(u_0+q)^{-\frac{4}{3}} , \quad \lambda(u_0)=\lambda_0(u_0+q)^{-\frac{4}{3}}\\
\end{split}
\label{i4}
\end{equation}
In addition, equation (\ref{i1}) is equivalent to the following integro-differential system of equations:
\begin{equation}
\begin{split}
 u_t-&v_x=0,\\
v_t-\Big{(}\int^{u}f(s)ds+&\varepsilon\lambda(u)v_x\Big{)}_x=0.
\end{split}
\label{i5}
\end{equation}
In the paper \cite{Ruggieri}, two different cases of equation (\ref{i5}) were considered:
\begin{equation}
\begin{split}
(IV):& \quad f(u_0)=f_0e^{\frac{1}{p}u_0} , \quad \lambda(u_0)=\lambda_0e^{\frac{1+s}{p}u_0}\\
(V):& \quad f(u_0)=f_0(u_0+q)^{\frac{1}{p}} , \quad \lambda(u_0)=\lambda_0(u_0+q)^{\frac{1+s}{p}-1}\\
\end{split}
\label{i6}
\end{equation}
and their Lie symmetry algebras were identified. The objectives of this work are the following. For each of the five cases listed in equations (\ref{i4}) and (\ref{i6}), we identify the classification of the one-dimensional subalgebras of the Lie symmetry algebra into conjugacy classes under the action of the associated Lie group. That is, we obtain a list of representative subalgebras of each Lie symmetry algebra ${\mathcal L}$ such that each one-dimensional subalgebra of ${\mathcal L}$ is conjugate to one and only one element of the list. In order to obtain these classifications, we make use of the results obtained by J. Patera and P. Winternitz in \cite{Patera}. For cases $(I)$ and $(II)$, we identify the Lie symmetry subalgebra as $2A_2$ from the list of Lie algebras of dimension $4$ found in \cite{Patera}. For case $(III)$, we first express the Lie symmetry subalgebra as a direct sum of two algebras, one of which is the three-dimensional algebra $A_{3,8}=su(1,1)$ found in \cite{Patera}. The Goursat method of twisted and non-twisted subalgebras is used to complete the classification \cite{Winternitz}. Next, we make a systematic use of the symmetry reduction method to generate invariant solutions corresponding to the above-mentioned subalgebras. We then perform a subalgebra classification for the integro-differential equation (\ref{i5}) and give two examples of symmetry reductions for this case. We provide a physical interpretation of the obtained results.

\section{Subalgebra classification and invariant solutions}

\subsection{The case where $f(u_0)=f_0e^{\frac{1}{p}u_0}$ and $\lambda(u_0)=\lambda_0e^{\frac{1+s}{p}u_0}$}

We first consider the case where $f(u_0)=f_0e^{\frac{1}{p}u_0}$ and $\lambda(u_0)=\lambda_0e^{\frac{1+s}{p}u_0}$, where $f_0$, $\lambda_0$, $p$ and $s$ are constants and $p\neq 0$. For this case, equations (\ref{bigeq1}) and (\ref{bigeq2}) become
\begin{equation}
u_{0,tt}-f_0e^{\frac{1}{p}u_0}u_{0,xx}-\dfrac{f_0}{p}e^{\frac{1}{p}u_0}(u_{0,x})^2=0,
\label{bigeq3}
\end{equation}
and 
\begin{equation}
\begin{split}
& u_{1,tt}-f_0e^{\frac{1}{p}u_0}u_{1,xx}-\dfrac{f_0}{p}e^{\frac{1}{p}u_0}u_{0,xx}u_1-\dfrac{2f_0}{p}e^{\frac{1}{p}u_0}u_{0,x}u_{1,x}-\dfrac{f_0}{p^2}e^{\frac{1}{p}u_0}(u_{0,x})^2u_1 \\ &-\lambda_0\left(\frac{1+s}{p}\right)^2e^{\frac{1+s}{p}u_0}(u_{0,x})^2u_{0,t}-\lambda_0\left(\frac{1+s}{p}\right)e^{\frac{1+s}{p}u_0}u_{0,xx}u_{0,t}\\ &-2\lambda_0\left(\frac{1+s}{p}\right)e^{\frac{1+s}{p}u_0}u_{0,x}u_{0,xt}-\lambda_0e^{\frac{1+s}{p}u_0}u_{0,xxt}=0.
\end{split}
\label{bigeq4}
\end{equation}
The Lie algebra of infinitesimal symmetries of equations (\ref{bigeq3}) and (\ref{bigeq4}) is spanned by the four generators \cite{Ruggieri}
\begin{equation}
\begin{split}
&X_1=\partial_t,\qquad X_2=\partial_x,\qquad X_3=t\partial_t+x\partial_x-u_1\partial_{u_1},\\  &X_4=x\partial_x+2p\partial_{u_0}+2su_1\partial_{u_1}.
\end{split}
\label{gen1}
\end{equation}
This Lie algebra is isomorphic to the algebra $2A_2$ given in Table II of \cite{Patera}. The list of conjugacy classes includes the following one-dimensional subalgebras:
\begin{equation}
\begin{split}
&\{X_1\}, \qquad \{X_4\}, \qquad \{X_2\}, \qquad \{X_3+aX_4\}, \qquad \{X_4-X_3+\varepsilon X_2\}, \\ &\{X_1+\varepsilon X_2\}, \qquad \{X_1+\varepsilon X_4\},
\end{split}
\label{subalg1}
\end{equation}
where $a\in \mathbb{R}$, $a\neq 0$ and $\varepsilon=\pm 1$. We proceed to use the symmetry reduction method to reduce the system of equations using each subalgebra given in the list (\ref{subalg1}).

{\bf 1.}\hspace{5mm} For the subalgebra $\{X_1\}$, we obtain the solution
\begin{equation}
\begin{split}
u_0(x)=p\ln{|x+C_1|}+C_2,\qquad u_1(x)=\dfrac{C_3}{x+C_1}+C_4,
\end{split}
\label{solution1}
\end{equation}
where $C_1$, $C_2$, $C_3$ and $C_4$ are constants. This is a singular logarithmic solution with one simple pole.

{\bf 2.}\hspace{5mm} For the subalgebra $\{X_4\}$, we obtain a dissipative solution of the form
\begin{equation}
\begin{split}
u_0(t,x)=F(t)+2p\ln{x},\qquad u_1(t,x)=x^{2s}G(t),
\end{split}
\label{solution2}
\end{equation}
where the functions $F(t)$ and $G(t)$ are given by the quadratures
\begin{equation}
\int\dfrac{dF}{\varepsilon(4p^2f_0e^{\frac{1}{p}F}+K_0)^{1/2}}=t-t_0,
\label{solution2e}
\end{equation}
for $F$ and 
\begin{equation}
G=\mu\int\sqrt{4b(s+1)(2s+1)\int e^{\frac{1}{p}F}[f_0G+\lambda_0\varepsilon e^{\frac{s}{p}F}(4p^2f_0e^{\frac{1}{p}F}+K_0)^{1/2}]}dFdt,
\label{solution2f}
\end{equation}
where $K_0$ and $b$ are constants and $\mu=\pm 1$. Therefore,
\begin{equation}
u_1=x^{2s}\mu\int\sqrt{4b(s+1)(2s+1)\int e^{\frac{1}{p}F}[f_0G+\lambda_0\varepsilon e^{\frac{s}{p}F}(4p^2f_0e^{\frac{1}{p}F}+K_0)^{1/2}]}dFdt.
\end{equation}

{\bf 3.}\hspace{5mm} For the subalgebra $\{X_2\}$, we obtain the trivial linear (in $t$) solution
\begin{equation}
u_0(t)=C_1t+C_2,\qquad u_1(t)=C_3t+C_4
\label{solution3}
\end{equation}
where $C_1$, $C_2$, $C_3$ and $C_4$ are constants.

{\bf 4.}\hspace{5mm} For the subalgebra $\{X_3+aX_4\}$, equations (\ref{bigeq3}) and (\ref{bigeq4}) reduce to the system of third-order ordinary differential equations (ODE)
\begin{equation}
(a+1)(a+2)\xi F_{\xi}+(a+1)^2\xi^2F_{\xi\xi}-2ap-f_0e^{\frac{1}{p}F}F_{\xi\xi}-\dfrac{f_0}{p}e^{\frac{1}{p}F}(F_{\xi})^2=0,
\label{bigeq3A}
\end{equation}
and
\begin{equation}
\begin{split}
&(2as-1)(2as-2)G-(a+1)(4as-a-4)\xi G_{\xi}+(a+1)^2\xi^2G_{\xi\xi}\\ &-f_0e^{\frac{1}{p}F}\left[G_{\xi\xi}+\dfrac{1}{p}F_{\xi\xi}G+\dfrac{2}{p}F_{\xi}G_{\xi}+\dfrac{1}{p^2}(F_{\xi})^2G\right]\\ &+\dfrac{\lambda_{0}(1+s)}{p}e^{\frac{1+s}{p}F}\Big{[}\dfrac{(a+1)(1+s)}{p}\xi(F_{\xi})^3-\dfrac{2ap(1+s)}{p}(F_{\xi})^2+(a+1)\xi F_{\xi}F_{\xi\xi}\\ &\hspace{3.5cm}-2apF_{\xi\xi}+2(a+1)(F_{\xi})^2+2(a+1)\xi F_{\xi}F_{\xi\xi}\Big{]}\\ &+\lambda_0e^{\frac{1+s}{p}F}\left[2(a+1)F_{\xi\xi}+(a+1)\xi F_{\xi\xi\xi}\right]=0,
\label{bigeq4A}
\end{split}
\end{equation}
where we have the self-similar symmetry variable $\xi=xt^{-a-1}$, and the functions
\begin{equation}
u_0=F(\xi)+2ap\ln{t}\qquad \mbox{and}\qquad u_1=t^{2as-1}G(\xi).
\end{equation}
For the special case of the subalgebra where $a=-1$, we obtain the singular logarithmic solution:
\begin{equation}
F(\xi)=2p\ln{\left(\frac{1}{\sqrt{f_0}}\xi+C_0\right)},
\label{solution4}
\end{equation}
where $C_0$ is a constant. The function $G$ satisfies the single second-order linear differential equation
\begin{equation}
\begin{split}
&-f_0\Delta^2G_{\xi\xi}-4\sqrt{f_0}\Delta G_{\xi}-2G+(2s+1)(2s+2)G\\ &+\dfrac{\lambda_0(1+s)\Delta^{2(1+s)}}{p}\left[\dfrac{4p^2}{\Delta^2}\left(\dfrac{2(1+s)}{f_0^2}-\dfrac{1}{f_0}\right)\right]=0,
\end{split}
\label{bigeq4AE}
\end{equation}
where $\Delta=\frac{1}{\sqrt{f_0}}\xi+C_0$. In the specific case where $\lambda_0=0$, we obtain the explicit solutions
\begin{equation}\label{earlydamp}
G=\xi^{-3/4}(C_1+C_2\ln{\xi})
\end{equation}
in the case where $s=-3/4$ and
\begin{equation}\label{otherearlydamp}
G=C_1\xi^{r_+}+C_2\xi^{r_-}
\end{equation}
where
\begin{equation}
r_{\pm}=\dfrac{-3\pm \sqrt{9-4[2-(2s+1)(2s+2)]}}{2}
\end{equation}
in the case where $s\neq -3/4$. The functions $G$ in equations (\ref{earlydamp}) and (\ref{otherearlydamp}) correspond respectively to the solutions
\begin{equation}\label{earlydampbis}
\begin{split}
& u_0=2p\ln{\left(\frac{1}{\sqrt{f_0}}xt^{-a-1}+C_0\right)}+2ap\ln{t},\\ & u_1=t^{2as-1}(xt^{-a-1})^{-3/4}(C_1+C_2\ln{(xt^{-a-1}}))
\end{split}
\end{equation}
and
\begin{equation}
\begin{split}
& u_0=2p\ln{\left(\frac{1}{\sqrt{f_0}}xt^{-a-1}+C_0\right)}+2ap\ln{t},\\ & u_1=t^{2as-1}(C_1(xt^{-a-1})^{r_+}+C_2(xt^{-a-1})^{r_-}).
\end{split}
\end{equation}
The solution (\ref{earlydampbis}) involves damping.

{\bf 5.}\hspace{5mm} For the subalgebra $\{X_4-X_3+\varepsilon X_2\}$, we get
\begin{equation}
u_0=F(\xi)-2p\ln{t},\qquad u_1=t^{-2s-1}G(\xi),
\label{solution5}
\end{equation}
where we have the symmetry variable $\xi=x+\varepsilon\ln{t}$. Here, $F$ satisfies the nonlinear equation
\begin{equation}
F_{\xi\xi}=\dfrac{1}{1-f_0e^{\frac{1}{p}F}}\left[\dfrac{f_0}{p}e^{\frac{1}{p}F}(F_{\xi})^2+\varepsilon F_{\xi}-2p\right],
\label{solution5A}
\end{equation}
and $G$ satisfies
\begin{equation}
\begin{split}
&\left(1-f_0e^{\frac{1}{p}F}\right)G_{\xi\xi}-\left(\varepsilon(4s+3)+\frac{2f_0}{p}e^{\frac{1}{p}F}F_{\xi}\right)G_{\xi}\\&+\left((2s+1)(2s+2)-\frac{f_0}{p}e^{\frac{1}{p}F}F_{\xi\xi}-\frac{f_0}{p^2}e^{\frac{1}{p}F}(F_{\xi})^2\right)G\\
&-\lambda_0\Bigg{[}\left(\dfrac{1+s}{p}\right)^2e^{\frac{1+s}{p}F}(F_{\xi})^2\left(\varepsilon F_{\xi}-2p\right)+\left(\dfrac{1+s}{p}\right)e^{\frac{1+s}{p}F}F_{\xi\xi}\left(\varepsilon F_{\xi}-2p\right)\\ &+2\varepsilon\left(\dfrac{1+s}{p}\right)e^{\frac{1+s}{p}F}F_{\xi}F_{\xi\xi}+\varepsilon e^{\frac{1+s}{p}F}F_{\xi\xi\xi} \Bigg{]}=0.
\end{split}
\end{equation}
In the specific case where $\lambda_0=0$ and $f_0=0$, we obtain the explicit solution
\begin{equation}\label{earlysolution1A}
u_0=K_1te^{\varepsilon x}+2\varepsilon px+K_2,\qquad\qquad u_1=K_3e^{(2s+1)x}t^{(2s+1)(\varepsilon-1)}+K_4e^{(2s+2)x}t^{(2s+1)(\varepsilon-1)}t^{\varepsilon}.
\end{equation}
Solution (\ref{earlysolution1A}) involves damping terms in the case when $\varepsilon=-1$. Otherwise, for $\varepsilon=1$, this solution may contain unbounded terms.

{\bf 6.}\hspace{5mm} For the subalgebra $\{X_1+\varepsilon X_2\}$, we have the travelling wave solution
\begin{equation}
u_0=u_0(\xi),\qquad u_1=u_1(\xi),
\label{solution6}
\end{equation}
where $\xi=x-\varepsilon t$. Here, $u_0$ can be determined implicitly by the equation
\begin{equation}
u_0-\dfrac{f_0}{p}e^{\frac{1}{p}u_0}=K_0\xi+K_1.
\label{solution6AAA}
\end{equation}
In the case where $\lambda_0=0$, $u_1$ satisfies the second-order ODE
\begin{equation}
\begin{split}
&\left(1-f_0e^{\frac{1}{p}u_0}\right)u_{1,\xi\xi}-\dfrac{2f_0}{p}e^{\frac{1}{p}u_0}\dfrac{K_0}{1-f_0e^{\frac{1}{p}u_0}}u_{1,\xi}\\ &-\dfrac{f_0}{p}e^{\frac{1}{p}u_0}\left[\dfrac{K_0f_0e^{\frac{1}{p}u_0}+K_0^2}{p\left(1-f_0e^{\frac{1}{p}u_0}\right)^2}\right]u_1=0
\end{split}
\label{solution6BBB}
\end{equation}
which is linear in $u_1$.

{\bf 7.}\hspace{5mm} For the subalgebra $\{X_1+\varepsilon X_4\}$, we obtain the center wave solution
\begin{equation}
u_0=F(\xi)+2\varepsilon pt,\qquad u_1=e^{2\varepsilon st}G(\xi),
\label{solution7}
\end{equation}
where the symmetry variable is $\xi=xe^{-\varepsilon t}$, $F$ satisfies the equation
\begin{equation}
\xi F_{\xi}+\xi^2F_{\xi\xi}-f_0e^{\frac{1}{p}F}F_{\xi\xi}-\dfrac{f_0}{p}e^{\frac{1}{p}F}(F_{\xi})^2=0,
\label{solution7A}
\end{equation}
and $G$ satisfies the equation
\begin{equation}
\begin{split}
&\left(\xi^2-f_0e^{\frac{1}{p}F}\right)G_{\xi\xi}+\left((1-4s)\xi-\dfrac{2f_0}{p}e^{\frac{1}{p}F}F_{\xi}\right)G_{\xi}\\ &+\left(4s^2-\dfrac{f_0}{p}e^{\frac{1}{p}F}F_{\xi\xi}-\dfrac{f_0}{p^2}e^{\frac{1}{p}F}(F_{\xi})^2\right)G\\ &+\lambda_0\varepsilon e^{\frac{1+s}{p}F}\bigg{[}\left(\dfrac{1+s}{p}\right)^2\xi(F_{\xi})^3-\dfrac{2s(1+s)}{p}(F_{\xi})^2+3\left(\dfrac{1+s}{p}\right)\xi F_{\xi}F_{\xi\xi}\\ &-2sFF_{\xi\xi}+\xi F_{\xi\xi\xi}\bigg{]}=0.
\end{split}
\label{solution7B}
\end{equation}
In the case where $\lambda_0=0$ and $s=\frac{1\pm\sqrt{2}}{2}$, we obtain the periodic damping solution
\begin{equation}
u_0=p\ln{x}+\varepsilon pt-\frac{p}{2}\ln{f_0},\qquad u_1=e^{2\varepsilon st+\frac{1}{2}xe^{-\varepsilon t}}\left[C_1\cos{\left(\frac{\sqrt{7}}{2}xe^{-\varepsilon t}\right)}+C_2\sin{\left(\frac{\sqrt{7}}{2}xe^{-\varepsilon t}\right)}\right].
\label{solution7C}
\end{equation}

\subsection{The case where $f(u_0)=f_0(u_0+q)^{\frac{1}{p}}$ and $\lambda(u_0)=\lambda_0(u_0+q)^{\frac{1+s}{p}-1}$}

Next, we consider the case where $f(u_0)=f_0(u_0+q)^{\frac{1}{p}}$ and $\lambda(u_0)=\lambda_0(u_0+q)^{\frac{1+s}{p}-1}$, where $f_0$, $\lambda_0$, $p$, $q$ and $s$ are constants with $p\neq 0$. For this case, equations (\ref{bigeq1}) and (\ref{bigeq2}) become
\begin{equation}
u_{0,tt}-f_0(u_0+q)^{\frac{1}{p}}u_{0,xx}-\dfrac{f_0}{p}(u_0+q)^{\frac{1}{p}-1}(u_{0,x})^2=0,
\label{bigeq5}
\end{equation}
and 
\begin{equation}
\begin{split}
& u_{1,tt}-f_0(u_0+q)^{\frac{1}{p}}u_{1,xx}-\dfrac{f_0}{p}(u_0+q)^{\frac{1}{p}-1}u_{0,xx}u_1-\dfrac{2f_0}{p}(u_0+q)^{\frac{1}{p}-1}u_{0,x}u_{1,x}\\ &-\dfrac{f_0}{p}\left(\dfrac{1}{p}-1\right)(u_0+q)^{\frac{1}{p}-2}(u_{0,x})^2u_1\\ &-\lambda_0\left(\dfrac{1+s}{p}-1\right)\left(\dfrac{1+s}{p}-2\right)(u_0+q)^{\frac{1+s}{p}-3}(u_{0,x})^2u_{0,t}\\ &-\lambda_0\left(\dfrac{1+s}{p}-1\right)(u_0+q)^{\frac{1+s}{p}-2}u_{0,xx}u_{0,t}\\ &-2\lambda_0\left(\dfrac{1+s}{p}-1\right)(u_0+q)^{\frac{1+s}{p}-2}u_{0,x}u_{0,xt}-\lambda_0(u_0+q)^{\frac{1+s}{p}-1}u_{0,xxt}=0.
\end{split}
\label{bigeq6}
\end{equation}
The Lie algebra of infinitesimal symmetries of equations (\ref{bigeq5}) and (\ref{bigeq6}) is spanned by the four generators \cite{Ruggieri}
\begin{equation}
\begin{split}
&X_1=\partial_t,\qquad X_2=\partial_x,\qquad X_3=t\partial_t+x\partial_x-u_1\partial_{u_1},\\  &X_4=x\partial_x+2p(u_0+q)\partial_{u_0}+2su_1\partial_{u_1}.
\end{split}
\label{gen2}
\end{equation}
This Lie algebra is isomorphic to the algebra $2A_2$ given in Table II of \cite{Patera}. The list of conjugacy classes includes the one-dimensional subalgebras:
\begin{equation}
\begin{split}
&\{X_1\}, \qquad \{X_4\}, \qquad \{X_2\}, \qquad \{X_3+aX_4\}, \qquad \{X_4-X_3+\varepsilon X_2\}, \\ &\{X_1+\varepsilon X_2\}, \qquad \{X_1+\varepsilon X_4\},
\end{split}
\label{subalg2}
\end{equation}
where $a\in \mathbb{R}$, $a\neq 0$ and $\varepsilon=\pm 1$. We obtain solutions of the equations by symmetry reduction using the different subalgebras in the list (\ref{subalg2}).

{\bf 8.}\hspace{5mm} For the subalgebra $\{X_1\}$, we obtain the explicit stationary solution
\begin{equation}
\begin{split}
u_0=\left(\dfrac{(p+1)(Kx+C)}{p}\right)^{\frac{p}{p+1}}-q,\qquad u_1=B_1(Kx+C)^{\frac{\sqrt{p}\lambda_1}{p+1}}+B_2(Kx+C)^{\frac{\sqrt{p}\lambda_2}{p+1}},
\end{split}
\label{solution8}
\end{equation}
where
\begin{equation}
\lambda=\dfrac{\dfrac{p-1}{\sqrt{p}}\pm\sqrt{\dfrac{(1-p)^2}{p}+4}}{2},
\label{solution8A}
\end{equation}
and $B_1$, $B_2$, $K$ and $C$ are constants. This solution involves a combination of powers of $x$.

{\bf 9.}\hspace{5mm} For the subalgebra $\{X_2\}$, we obtain the trivial linear (in $t$) solution
\begin{equation}
u_0=C_1t+C_2,\qquad u_1=C_3t+C_4,
\label{solution9}
\end{equation}
where $C_1$, $C_2$, $C_3$ and $C_4$ are constants.

{\bf 10.}\hspace{5mm} For the subalgebra $\{X_4\}$, we obtain
\begin{equation}
\begin{split}
u_0=x^{2p}F(t)-q,\qquad u_1=x^{2s}G(t),
\end{split}
\label{solution10}
\end{equation}
where
\begin{equation}
F=\left(\varepsilon\sqrt{f_0}(t-t_0)\right)^{-2p},
\label{solution10e}
\end{equation}
and $G$ satisfies the linear second-order ODE
\begin{equation}
G_{tt}-f_0(4s^2+6s+2)(\varepsilon\sqrt{f_0}(t-t_0))^{-2}G+2\lambda_0\varepsilon\sqrt{f_0}p(4s^2+6s+2)(\varepsilon\sqrt{f_0}(t-t_0))^{-2s-3}=0.
\label{solution10f}
\end{equation}
The function $F$ involves damping if $p>0$. In the specific case where $\lambda_0=0$, $t_0=0$ and either $s=-2$ or $s=\frac{1}{2}$, we obtain $G=C_1t^3+C_2t^{-2}$, so the solution is
\begin{equation}
u_0=x^{2p}\left(\varepsilon\sqrt{f_0}(t-t_0)\right)^{-2p}-q,\qquad u_1=x^{2s}\left(C_1t^3+C_2t^{-2}\right).
\end{equation}
In the specific case where $\lambda_0=0$, $t_0=0$ and either $s=1$ or $s=-\frac{5}{2}$, we obtain $G=C_1t^4+C_2t^{-3}$, so the solution is
\begin{equation}
u_0=x^{2p}\left(\varepsilon\sqrt{f_0}(t-t_0)\right)^{-2p}-q,\qquad u_1=x^{2s}\left(C_1t^4+C_2t^{-3}\right).
\end{equation}
These solutions involve combinations of powers of $x$ and $t$.

{\bf 11.}\hspace{5mm} For the subalgebra $\{X_3+aX_4\}$, we get
\begin{equation}
\begin{split}
u_0=t^{2ap}F(\xi)-q,\qquad u_1=t^{2as-1}G(\xi),
\end{split}
\label{solution11}
\end{equation}
where the self-similar invariant has the form $\xi=xt^{-a-1}$, with $F=\dfrac{(a+1)^{2p}}{f_0^p}\xi^{2p}$ and $G=R\xi^{2s}$, where $R$ is a constant. Here, the following conditions have to be satisfied:
\begin{equation}
\begin{split}
&\mbox{(1) }\quad a(a+2)(2p+1)=0\\
&\mbox{(2) }\quad -2a(2as^2+4s^2+3as+6s+a+2)+4\lambda_0p(1+3s+2s^2)=0
\end{split}
\label{conditions11AE}
\end{equation}
Equation (\ref{solution11}) leads to a power function solution.

{\bf 12.}\hspace{5mm} For the subalgebra $\{X_4-X_3+\varepsilon X_2\}$, we get
\begin{equation}
u_0=t^{-2p}F(\xi)-q,\qquad u_1=t^{-2s-1}G(\xi),
\label{solution12}
\end{equation}
with symmetry variable $\xi=x+\varepsilon\ln{t}$. Here, $F$ satisfies the equation
\begin{equation}
\left(1-f_0F^{\frac{1}{p}}\right)F_{\xi\xi}-\dfrac{f_0}{p}F^{\frac{1}{p}-1}(F_{\xi})^2-\varepsilon(4p+1)F_{\xi}+2p(2p+1)F=0.
\label{solution12A}
\end{equation}
In the case where $p=-\frac{1}{2}$, we obtain the implicitly-defined function
\begin{equation}
-\varepsilon\ln{(A-\varepsilon F)}+\dfrac{f_0}{A^2}\left(\dfrac{A-\varepsilon F}{F}+\varepsilon\ln{\left(\dfrac{A-\varepsilon F}{F}\right)}\right)=\xi-\xi_0.
\label{solution12B}
\end{equation}
The equation for $G(\xi)$ in this case becomes
\begin{equation}
\begin{split}
&\left(1-f_0F^{-2}\right)G_{\xi\xi}+\left(-4s\varepsilon-3\varepsilon+4f_0F^{-3}F_{\xi}\right)G_{\xi}\\ &+\left((2s+1)(2s+2)+2f_0F^{-3}F_{\xi\xi}-6f_0F^{-4}(F_{\xi})^2\right)G\\ &-\lambda_0(2s+3)(2s+4)F^{-2s-5}(F_{\xi})^2\left[F+\varepsilon F_{\xi}\right]+\lambda_0(2s+3)F^{-2s-4}F_{\xi\xi}\left[F+\varepsilon F_{\xi}\right]\\ &+2\lambda_0(2s+3)F^{-2s-4}F_{\xi}\left[F_{\xi}+\varepsilon F_{\xi\xi}\right]-\lambda_0F^{-2s-3}\left[F_{\xi\xi}+\varepsilon F_{\xi\xi\xi}\right]=0.
\end{split}
\label{solution12C}
\end{equation}
If we further suppose that $\lambda_0=0$ and $f_0=0$, we obtain the solution
\begin{equation}
u_0=\varepsilon At^{-2p}-\varepsilon t^{-2p-1}e^{-\varepsilon x}e^{\varepsilon\xi_0}-q,\qquad u_1=C_1e^{\lambda_1x}t^{\varepsilon\lambda_1-2s-1}+C_2e^{\lambda_2x}t^{\varepsilon\lambda_2-2s-1},
\end{equation}
where
\begin{equation}
\lambda_1=\dfrac{4\varepsilon s+3\varepsilon+1}{2},\qquad \lambda_2=\dfrac{4\varepsilon s+3\varepsilon-1}{2}.
\end{equation}

{\bf 13.}\hspace{5mm} For the subalgebra $\{X_1+\varepsilon X_2\}$, we have the travelling wave solution
\begin{equation}
u_0=u_0(\xi),\qquad u_1=u_1(\xi),
\label{solution13}
\end{equation}
where $\xi=x-\varepsilon t$ is the symmetry variable. Here, $u_0$ satisfies
\begin{equation}
\left(1-f_0(u_0+q)^{\frac{1}{p}}­\right)u_{0,\xi\xi}=\dfrac{f_0}{p}(u_0+q)^{\frac{1}{p}-1}(u_{0,\xi})^2,
\label{solution13A}
\end{equation}
and $u_1$ satisfies
\begin{equation}
\begin{split}
&\left(1-f_0(u_0+q)^{\frac{1}{p}}\right)u_{1,\xi\xi}-\dfrac{2f_0}{p}(u_0+q)^{\frac{1}{p}-1}u_{0,\xi}u_{1,\xi}\\ &-\dfrac{f_0}{p}\left[(u_0+q)^{\frac{1}{p}-1}u_{0,\xi\xi}+\left(\dfrac{1}{p}-1\right)(u_0+q)^{\frac{1}{p}-2}(u_{0,\xi})^2\right]u_1\\ &+\varepsilon\lambda_0\bigg{[}\left(\dfrac{1+s}{p}-1\right)\left(\dfrac{1+s}{p}-2\right)(u_0+q)^{\frac{1+s}{p}-3}(u_{0,\xi})^3\\ &+3\left(\dfrac{1+s}{p}-1\right)(u_0+q)^{\frac{1+s}{p}-2}u_{0,\xi}u_{0,\xi\xi}+(u_0+q)^{\frac{1+s}{p}-1}u_{0,\xi\xi\xi}\bigg{]}=0.
\end{split}
\label{solution13B}
\end{equation}
In the case where $\lambda_0=0$, we obtain the explicit solution
\begin{equation}
u_0=\dfrac{1}{(f_0)^p}-q,\quad\mbox{while}\quad u_1=u_1(\xi)\quad\mbox{is an arbitrary function of }\xi.
\end{equation}

{\bf 14.}\hspace{5mm} For the subalgebra $\{X_1+\varepsilon X_4\}$, we obtain
\begin{equation}
u_0=x^{2p}F(\xi)-q,\qquad u_1=x^{2s}G(\xi),
\label{solution14}
\end{equation}
where the symmetry variable is $\xi=\ln{x}-\varepsilon t$ and $F$ satisfies the equation
\begin{equation}
\left(1-f_0F^{\frac{1}{p}}\right)F_{\xi\xi}-\dfrac{f_0}{p}F^{\frac{1}{p}-1}(F_{\xi})^2-f_0F^{\frac{1}{p}}(4p+3)F_{\xi}-2p(2p+1)f_0F^{\frac{1}{p}+1}=0,
\label{solution14A}
\end{equation}
and $G$ satisfies
\begin{equation}
\begin{split}
& \left(1-f_0F^{\frac{1}{p}}\right)G_{\xi\xi}-\left(f_0(4s-1)F^{\frac{1}{p}}+\dfrac{2f_0}{p}F^{\frac{1}{p}-1}\left(2pF+F_{\xi}\right)\right)G_{\xi}\\ & -\Bigg{(}2s(2s-1)f_0F^{\frac{1}{p}}+\dfrac{4sf_0}{p}F^{\frac{1}{p}-1}\left(2pF+F_{\xi}\right)\\ & +\dfrac{f_0}{p}F^{\frac{1}{p}-1}\left[2p(2p-1)F+(4p-1)F_{\xi}+F_{\xi\xi}\right]\\ & +\dfrac{f_0}{p}\left(\dfrac{1}{p}-1\right)F^{\frac{1}{p}-2}\left[4p^2F^2+4pFF_{\xi}+(F_{\xi})^2\right]\Bigg{)}G\\ & +\varepsilon\lambda_0\Bigg{[}\left(\dfrac{1+s}{p}-1\right)\left(\dfrac{1+s}{p}-2\right)F^{\frac{1+s}{p}-3}F_{\xi}\left(4p^2F^2+4pFF_{\xi}+(F_{\xi})^2\right)\\ & +\left(\dfrac{1+s}{p}-1\right)F^{\frac{1+s}{p}-2}F_{\xi}\left(2p(2p-1)F+(4p-1)F_{\xi}+F_{\xi\xi}\right)\\ & +2\left(\dfrac{1+s}{p}-1\right)F^{\frac{1+s}{p}-2}\left(2pF+F_{\xi}\right)\left(2pF_{\xi}+F_{\xi\xi}\right)\\ & +F^{\frac{1+s}{p}-1}\left(2p(2p-1)F_{\xi}+(4p-1)F_{\xi\xi}+F_{\xi\xi\xi}\right)\Bigg{]}=0.
\end{split}
\end{equation}
In the case where $p=-\frac{1}{2}$ and $\lambda_0=0$, we obtain the solution
\begin{equation}
u_0=x^{2p}\sqrt{f_0}-q,\qquad\qquad u_1=K_0x^{2s-r}e^{\varepsilon rt},\quad\mbox{where}\quad r=\dfrac{4s^2+6s+2}{4s+3}.
\end{equation}

\subsection{The case where $f(u_0)=f_0(u_0+q)^{-\frac{4}{3}}$ and $\lambda(u_0)=\lambda_0(u_0+q)^{-\frac{4}{3}}$}

We now consider the case where $f(u_0)=f_0(u_0+q)^{-\frac{4}{3}}$ and $\lambda(u_0)=\lambda_0(u_0+q)^{-\frac{4}{3}}$, where $f_0$, $\lambda_0$ and $q$ are constants. This corresponds to the special instance of the previous case (in subsection 2.2) in which $p=-\frac{3}{4}$ and $s=-\frac{3}{4}$. For this case, equations (\ref{bigeq1}) and (\ref{bigeq2}) become
\begin{equation}
u_{0,tt}-f_0(u_0+q)^{-\frac{4}{3}}u_{0,xx}+\dfrac{4}{3}f_0(u_0+q)^{-\frac{7}{3}}(u_{0,x})^2=0,
\label{bigeq7}
\end{equation}
and 
\begin{equation}
\begin{split}
& u_{1,tt}-f_0(u_0+q)^{-\frac{4}{3}}u_{1,xx}+\dfrac{4}{3}f_0(u_0+q)^{-\frac{7}{3}}u_{0,xx}u_1+\dfrac{8}{3}f_0(u_0+q)^{-\frac{7}{3}}u_{0,x}u_{1,x}\\ &-\dfrac{28}{9}f_0(u_0+q)^{-\frac{10}{3}}(u_{0,x})^2u_1-\dfrac{28}{9}\lambda_0(u_0+q)^{-\frac{10}{3}}(u_{0,x})^2u_{0,t}\\ &+\dfrac{4}{3}\lambda_0(u_0+q)^{-\frac{7}{3}}u_{0,xx}u_{0,t}+\dfrac{8}{3}\lambda_0(u_0+q)^{-\frac{7}{3}}u_{0,x}u_{0,xt}-\lambda_0(u_0+q)^{-\frac{4}{3}}u_{0,xxt}=0.
\end{split}
\label{bigeq8}
\end{equation}
The Lie algebra of infinitesimal symmetries of equations (\ref{bigeq7}) and (\ref{bigeq8}) is spanned by the five generators \cite{Ruggieri}
\begin{equation}
\begin{split}
&X_1=\partial_t,\qquad X_2=\partial_x,\qquad X_3=t\partial_t+x\partial_x-u_1\partial_{u_1},\\  &X_4=x\partial_x-\dfrac{3}{2}(u_0+q)\partial_{u_0}-\dfrac{3}{2}u_1\partial_{u_1},\qquad X_5=x^2\partial_x-3x(u_0+q)\partial_{u_0}-3xu_1\partial_{u_1}.
\end{split}
\label{gen3}
\end{equation}
This Lie algebra is the direct sum
\begin{equation}
\{X_3-X_4,X_1\}\oplus\{X_4,X_2,X_5\},
\end{equation}
where $\{X_4,X_2,X_5\}$ is isomorphic to the three-dimensional algebra $A_{3,8}=su(1,1)$ given in Table I of \cite{Patera}. The classification of $A_{3,8}$ was found in \cite{Patera} and, in this paper, the Goursat method of twisted and non-twisted subalgebras is used to obtain the list of conjugacy classes for the complete Lie symmetry algebra. The one-dimensional subalgebras of the Lie algebra can be classified as follows:
\begin{equation}
\begin{split}
&\{X_3-X_4\}, \qquad \{X_1\}, \qquad \{X_2\}, \qquad \{X_4\}, \qquad \{X_2-X_5\},\\ & \{X_3-X_4+\varepsilon X_2\}, \qquad \{X_3+aX_4\}, \qquad \{X_3-X_4+a(X_2-X_5)\},\\ & \{X_1+\varepsilon X_2\}, \qquad \{X_1+\varepsilon X_4\}, \qquad \{X_1+\varepsilon(X_2-X_5)\},
\end{split}
\label{subalg3}
\end{equation}
where $a\in \mathbb{R}$, $a\neq 0$ and $\varepsilon=\pm 1$. We obtain the following solutions through symmetry reduction.

{\bf 15.}\hspace{5mm} For the subalgebra $\{X_3-X_4\}$, we obtain the power function solution
\begin{equation}
\begin{split}
u_0=\left(\dfrac{t}{x}\right)^{\frac{3}{2}}-q,\qquad u_1=\dfrac{t^{\frac{1}{2}}}{x^{\frac{3}{2}}},
\end{split}
\label{solution15}
\end{equation}
for the case where $f_0=1$ and $\lambda_0=0$. A second solution, obtained by making the hypothesis $F=C_0x^a$, is
\begin{equation}
u_0=f_0^{\frac{3}{4}}\left(\dfrac{t}{x}\right)^{\frac{3}{2}}-q,\qquad u_1=t^{\frac{1}{2}}\left[C_1x^{\frac{-3+\sqrt{\frac{140}{3}}}{2}}+C_2x^{\frac{-3-\sqrt{\frac{140}{3}}}{2}}-\dfrac{69}{280f_0^{\frac{1}{4}}}x^{-\frac{3}{2}}\right],
\label{solution15A}
\end{equation}
which constitutes a combination of monomial power functions.

{\bf 16.}\hspace{5mm} For the subalgebra $\{X_4\}$, we get the center wave solution
\begin{equation}
\begin{split}
&u_0=\dfrac{f_0^{\frac{3}{4}}t^{\frac{3}{2}}}{x^{\frac{3}{2}}}-q,\\
&u_1=\dfrac{C_1t^{\frac{1}{2}}}{x^{\frac{3}{2}}}+\dfrac{C_2t^{\frac{1}{2}}\ln{t}}{x^{\frac{3}{2}}}+\dfrac{3\lambda_0}{32f_0^{\frac{1}{4}}}\dfrac{t^{\frac{5}{2}}}{x^{\frac{3}{2}}}(2\ln{t}-1)-\dfrac{3\lambda_0}{16f_0^{\frac{1}{4}}}\dfrac{t^{\frac{1}{2}}}{x^{\frac{3}{2}}}\ln{t}.
\end{split}
\label{solution16}
\end{equation}

{\bf 17.}\hspace{5mm} For the subalgebra $\{X_2\}$, we obtain the linear trivial solution in $t$
\begin{equation}
u_0=C_1t+C_2,\qquad u_1=C_3t+C_4,
\label{solution17}
\end{equation}
where $C_1$, $C_2$, $C_3$ and $C_4$ are constants.

{\bf 18.}\hspace{5mm} For the subalgebra $\{X_1\}$, we have $u_0=u_0(x)$ and $u_1=u_1(x)$ (i.e. $u_0$ and $u_1$ are functions of $x$ only), where $u_0$ satisfies the equation
\begin{equation}
u_{0,xx}=\dfrac{4(u_{0,x})^2}{3(u_0+q)},
\label{solution18}
\end{equation}
and $u_1$ satisfies the equation
\begin{equation}
\begin{split}
u_{1,xx}=\dfrac{4}{3(u_0+q)}u_{0,xx}u_1+\dfrac{8}{3(u_0+q)}u_{0,x}u_{1,x}-\dfrac{28}{9(u_0+q)^2}(u_{0,x})^2u_1.
\end{split}
\label{solution18A}
\end{equation}
For the specific case when $q=0$, the solution of equation (\ref{solution18}) is expressed in terms of the Gaussian quadrature
\begin{equation}
\int e^{-\frac{2}{3}u_0^2}du_0=k(x-x_0),
\label{solution18B}
\end{equation}
and equation (\ref{solution18A}) becomes the second-order ordinary differential equation
\begin{equation}
u_{1,xx}=\dfrac{4k}{3u_0}e^{\frac{2}{3}u_0^2}\left(2u_{1,x}-\dfrac{1}{u_0}ke^{\frac{2}{3}u_0^2}u_1\right),
\label{solution18C}
\end{equation}
where $k$ is a constant.

{\bf 19.}\hspace{5mm} For the subalgebra $\{X_2-X_5\}$, we get
\begin{equation}
\begin{split}
u_0=(1-x^2)^{-\frac{3}{2}}F(t)-q,\qquad u_1=(1-x^2)^{-\frac{3}{2}}G(t),
\end{split}
\label{solution19}
\end{equation}
where the functions $F$ and $G$ of $t$ satisfy the equations
\begin{equation}
F_{tt}-3f_0F^{-\frac{1}{3}}=0,
\label{solution19A}
\end{equation}
and
\begin{equation}
G_{tt}+f_0F^{-\frac{4}{3}}G+\lambda_0 F^{-\frac{4}{3}}F_t=0.
\label{solution19B}
\end{equation}
In the case where $\lambda_0=0$, looking for solutions of the type $F=At^a$, $G=Bt^b$, we obtain the solution
\begin{equation}
u_0=(1-x^2)^{-\frac{3}{2}}(4f_0)^{3/4}t^{3/2}-q,\qquad u_1=(1-x^2)^{-\frac{3}{2}}Bt^{1/2},\qquad \mbox{where }B\mbox{ is a constant}.
\end{equation}
This solution involves a separation of the variables $x$ and $t$.

{\bf 20.}\hspace{5mm} For the subalgebra $\{X_3-X_4+\varepsilon X_2\}$, we get
\begin{equation}
u_0=t^{\frac{3}{2}}F(\xi)-q,\qquad u_1=t^{\frac{1}{2}}G(\xi),
\label{solution20}
\end{equation}
where the functions $F$ and $G$ of the symmetry variable $\xi=x-\varepsilon\ln{t}$ satisfy the equations
\begin{equation}
\left(1-f_0F^{-\frac{4}{3}}\right)F_{\xi\xi}+\dfrac{4}{3}f_0F^{-\frac{7}{3}}(F_{\xi})^2-2\varepsilon F_{\xi}+\dfrac{3}{4}F=0,
\label{solution20A}
\end{equation}
and
\begin{equation}
\begin{split}
&\left(1-f_0F^{-\frac{4}{3}}\right)G_{\xi\xi}+\dfrac{8}{3}f_0F^{-\frac{7}{3}}F_{\xi}G_{\xi}\\ &+\left(\dfrac{4}{3}f_0F^{-\frac{7}{3}}F_{\xi\xi}-\dfrac{28}{9}f_0F^{-\frac{10}{3}}(F_{\xi})^2-\dfrac{1}{4}\right)G\\ &+\lambda_0F^{-\frac{10}{3}}\bigg{(}-\dfrac{2}{3}F(F_{\xi})^2+\dfrac{28}{9}\varepsilon(F_{\xi})^3+\dfrac{1}{2}F^2F_{\xi\xi}-4\varepsilon FF_{\xi}F_{\xi\xi}+\varepsilon F^2F_{\xi\xi\xi}\bigg{)}=0.
\end{split}
\label{solution20B}
\end{equation}
Here, $F(\xi)$ is the function such that
\begin{equation}
\left(\eta+2\varepsilon F\right)\eta'=1,
\end{equation}
where $F$ and $\eta$ obey the constraints
\begin{equation}
\eta=\eta(\zeta)=F_{\xi}\left(1-f_0F^{-4/3}\right)-2\varepsilon F,\qquad\mbox{and}\qquad \zeta=-\frac{3}{8}F^2+\frac{9}{8}f_0F^{2/3}.
\end{equation}

{\bf 21.}\hspace{5mm} For the subalgebra $\{X_3+aX_4\}$, we obtain
\begin{equation}
\begin{split}
u_0=t^{-\frac{3a}{2}}F(\xi)-q,\qquad u_1=t^{-\frac{3a+2}{2}}G(\xi),
\end{split}
\label{solution21}
\end{equation}
where $F$ and $G$ are functions of the self-similar symmetry variable $\xi=xt^{-a-1}$. Here, $F$ satisfies the equation
\begin{equation}
\begin{split}
&\left((a+1)^2\xi^2-f_0F^{-\frac{4}{3}}\right)F_{\xi\xi}+\dfrac{4}{3}f_0F^{-\frac{7}{3}}(F_{\xi})^2+2(a+1)(2a+1)\xi F_{\xi}\\ &+\dfrac{3a(3a+2)}{4}F=0.
\end{split}
\label{solution21A}
\end{equation}
In the case where $\lambda_0=0$ and either $a=0$ or $a=-2$, the function
\begin{equation}
F=f_0^{\frac{3}{4}}(a+1)^{-\frac{3}{2}}\xi^{-\frac{3}{2}}
\label{solution21AAA}
\end{equation}
is a solution with damping of equation (\ref{solution21A}). Substituting the function (\ref{solution21AAA}) and any arbitrary function $G(\xi)$ of the symmetry variable $\xi=xt^{-a-1}$ into (\ref{solution21}), we obtain a solution of the system consisting of equations (\ref{bigeq7}) and (\ref{bigeq8}) of the form
\begin{equation}
u_0=f_0^{\frac{3}{4}}(a+1)^{-\frac{3}{2}}x^{-\frac{3}{2}}t^{\frac{3}{2}}-q,\qquad u_1=t^{-\frac{3a+2}{2}}G(\xi),
\end{equation}
 where $G$ is an arbitrary function of $\xi=xt^{-a-1}$.

{\bf 22.}\hspace{5mm} For the subalgebra $\{X_1+\varepsilon X_2\}$, we obtain the travelling wave solution
\begin{equation}
u_0=u_0(\xi),\qquad u_1=u_1(\xi),
\label{solution22}
\end{equation}
where we have $\xi=x-\varepsilon t$. Here, $u_0$ satisfies the equation
\begin{equation}
\left(1-f_0(u_0+q)^{-\frac{4}{3}}­\right)u_{0,\xi\xi}+\dfrac{4}{3}f_0(u_0+q)^{-\frac{7}{3}}(u_{0,\xi})^2=0,
\label{solution22A}
\end{equation}
and $u_1$ satisfies
\begin{equation}
\begin{split}
&\left(1-f_0(u_0+q)^{-\frac{4}{3}}\right)u_{1,\xi\xi}+\dfrac{8}{3}f_0(u_0+q)^{-\frac{7}{3}}u_{0,\xi}u_{1,\xi}\\ &+f_0\left(\dfrac{4}{3}(u_0+q)^{-\frac{7}{3}}u_{0,\xi\xi}-\dfrac{28}{9}(u_0+q)^{-\frac{10}{3}}(u_{0,\xi})^2\right)u_1\\ &+\lambda_0\bigg{(}\dfrac{28}{9}\varepsilon(u_0+q)^{-\frac{10}{3}}(u_{0,\xi})^3-4\varepsilon(u_0+q)^{-\frac{7}{3}}u_{0,\xi}u_{0,\xi\xi}+\varepsilon(u_0+q)^{-\frac{4}{3}}u_{0,\xi\xi\xi}\bigg{)}=0.
\end{split}
\label{solution22B}
\end{equation}
Equation (\ref{solution22A}) can be solved implicitly through the quadrature
\begin{equation}
\int\dfrac{du_0}{\ln{\left(1-f_0(u_0+q)^{-\frac{4}{3}}\right)}}=\xi_0-\xi.
\end{equation}

{\bf 23.}\hspace{5mm} For the subalgebra $\{X_1+\varepsilon X_4\}$, we get
\begin{equation}
u_0=x^{-\frac{3}{2}}F(\xi)-q,\qquad u_1=x^{-\frac{3}{2}}G(\xi),
\label{solution23}
\end{equation}
where we have the symmetry variable $\xi=t-\varepsilon\ln{x}$, $x>0$. Here, $F$ and $G$ satisfy the equations
\begin{equation}
\left(1-f_0F^{-\frac{4}{3}}\right)F_{\xi\xi}+\dfrac{4}{3}f_0F^{-\frac{7}{3}}(F_{\xi})^2-\dfrac{3}{4}f_0F^{-\frac{1}{3}}=0,
\label{solution23A}
\end{equation}
and
\begin{equation}
\begin{split}
&\left(1-f_0F^{-\frac{4}{3}}\right)G_{\xi\xi}+\dfrac{8}{3}f_0F^{-\frac{7}{3}}F_{\xi}G_{\xi}\\ &+\left(\dfrac{4}{3}f_0F^{-\frac{7}{3}}F_{\xi\xi}-\dfrac{28}{9}f_0F^{-\frac{10}{3}}(F_{\xi})^2+\dfrac{1}{4}f_0F^{-\frac{4}{3}}\right)G\\ &+\lambda_0\bigg{(}\dfrac{1}{4}F^{-\frac{4}{3}}F_{\xi}-\dfrac{28}{9}F^{-\frac{10}{3}}(F_{\xi})^3+4F^{-\frac{7}{3}}F_{\xi}F_{\xi\xi}-F^{-\frac{4}{3}}F_{\xi\xi\xi}\bigg{)}=0,
\end{split}
\label{solution23B}
\end{equation}
respectively.
Equation (\ref{solution23A}) can be solved implicitly through the quadrature
\begin{equation}
\int\dfrac{f_0F^{-\frac{4}{3}}-1}{\sqrt{\frac{9}{4}(f_0)^2F^{-\frac{2}{3}}+\frac{9}{4}f_0F^{\frac{2}{3}}+K}}dF=\xi-\xi_0.
\end{equation}

{\bf 24.}\hspace{5mm} For the subalgebra $\{X_1+\varepsilon(X_2-X_5)\}$, we have
\begin{equation}
u_0=(x^2-1)^{-\frac{3}{2}}F(\xi)-q,\qquad u_1=(x^2-1)^{-\frac{3}{2}}G(\xi),
\label{solution24}
\end{equation}
where we have the symmetry variable
\begin{equation}
\xi=\varepsilon t+\frac{1}{2}\ln{\left(\frac{x-1}{x+1}\right)}.
\end{equation}
Here, $F$ and $G$ satisfy the equations
\begin{equation}
\left(1-f_0F^{-\frac{4}{3}}\right)F_{\xi\xi}+\dfrac{4}{3}f_0F^{-\frac{7}{3}}(F_{\xi})^2-3f_0F^{-\frac{1}{3}}=0,
\label{solution24A}
\end{equation}
and
\begin{equation}
\begin{split}
&\left(1-f_0F^{-\frac{4}{3}}\right)G_{\xi\xi}+\dfrac{8}{3}f_0F^{-\frac{7}{3}}F_{\xi}G_{\xi}\\ &+\left(\dfrac{4}{3}f_0F^{-\frac{7}{3}}F_{\xi\xi}-\dfrac{28}{9}f_0F^{-\frac{10}{3}}(F_{\xi})^2+f_0F^{-\frac{4}{3}}\right)G\\ &+\lambda_0\varepsilon\bigg{(}F^{-\frac{4}{3}}F_{\xi}-\dfrac{28}{9}F^{-\frac{10}{3}}(F_{\xi})^3+4F^{-\frac{7}{3}}F_{\xi}F_{\xi\xi}-F^{-\frac{4}{3}}F_{\xi\xi\xi}\bigg{)}=0.
\end{split}
\label{solution24B}
\end{equation}
Equation (\ref{solution24A}) can be solved implicitly through the quadrature
\begin{equation}
\int\dfrac{f_0F^{-\frac{4}{3}}-1}{\sqrt{9(f_0)^2F^{-\frac{2}{3}}+9f_0F^{\frac{2}{3}}+K}}dF=\xi-\xi_0.
\end{equation}

{\bf 25.}\hspace{5mm} For the subalgebra $\{X_3-X_4+a(X_2-X_5)\}$, we consider the case where $a=\frac{1}{2}$. We obtain
\begin{equation}
u_0=(x-1)^{-3}F(\xi)-q,\qquad u_1=(x-1)^{-2}(x+1)^{-1}G(\xi),
\label{solution25}
\end{equation}
where the rational symmetry variable is $\xi=t(x-1)(x+1)^{-1}$. Here, $F$ satisfies the equation
\begin{equation}
\left(1-4f_0\xi^2F^{-\frac{4}{3}}\right)F_{\xi\xi}+\dfrac{16}{3}f_0\xi^2F^{-\frac{7}{3}}(F_{\xi})^2-8f_0\xi F^{-\frac{4}{3}}F_{\xi}=0.
\label{solution25A}
\end{equation}
A particular solution is
\begin{equation}
F=2^{\frac{3}{2}}f_0^{\frac{3}{4}}\xi^{\frac{3}{2}}.
\label{solution25B}
\end{equation}
In the case where $\lambda_0=0$ and $a=\frac{1}{2}$, substituting the function (\ref{solution25B}) and any arbitrary function $G(\xi)$ of the symmetry variable $\xi=t(x-1)(x+1)^{-1}$ into (\ref{solution25}) yields a solution of the system consisting of equations (\ref{bigeq7}) and (\ref{bigeq8})
\begin{equation}
u_0=2^{\frac{3}{2}}f_0^{\frac{3}{4}}t^{\frac{3}{2}}(x-1)^{-\frac{3}{2}}(x+1)^{-\frac{3}{2}}-q.
\end{equation}

\section{Subalgebra classification and solutions for the integro-differential case}

The system (\ref{i5}) given by the equations
\begin{equation}
\begin{split}
 u_t-&v_x=0\\
v_t-\big{(}\int^{u}f(s)ds+&\varepsilon\lambda(u)v_x\big{)}_x=0 
\end{split}
\label{integralformA}
\end{equation}
is the potential system for equation (\ref{i2}) in the sense that its compatibility condition is given by equation (\ref{i2}). Here, we have
\begin{equation}
u(t,x,\varepsilon)=u_0(t,x)+\varepsilon u_1(t,x)+{\mathcal O}(\varepsilon^2)\quad\mbox{ and }\quad v(t,x,\varepsilon)=v_0(t,x)+\varepsilon v_1(t,x)+{\mathcal O}(\varepsilon^2)
\label{uandvform}
\end{equation}
The approximate Lie algebra of infinitesimal symmetries of equation (\ref{integralformA}) is spanned by the five generators \cite{Ruggieri}
\begin{equation}
\begin{split}
&X_1=\partial_t,\qquad X_2=\partial_x,\qquad X_3=\partial_{v_0},\qquad X_4=\partial_{v_1},\\  &X_5=t\partial_t+x\partial_x-u_1\partial_{u_1}-v_1\partial_{v_1}
\end{split}
\label{gen4}
\end{equation}
For two specific cases of $f(u_0)$ and $\lambda(u_0)$, we also have an additional generator $X_6$. Specifically:
\begin{itemize}
\item For the case where $f(u_0)=f_0e^{u_0/p}$ and $\lambda(u_0)=\lambda_0e^{(1+s)u_0/p}$, we have\\ $X_6=x\partial_x+2p\partial_{u_0}+v_0\partial_{v_0}+2su_1\partial_{u_1}+(2s+1)v_1\partial_{v_1}$
\item For the case where $f(u_0)=f_0(u_0+q)^{\frac{1}{p}}$ and $\lambda(u_0)=\lambda_0(u_0+q)^{\frac{1+s}{p}-1}$, we have $X_6=x\partial_x+2p(u_0+q)\partial_{u_0}+(2p+1)v_0\partial_{v_0}+2su_1\partial_{u_1}+(2s+1)v_1\partial_{v_1}$
\end{itemize}
For both cases, we obtain a classification of 63 conjugacy classes of one-dimensional subalgebras, which we list in the Appendix.

\subsection{The case where $f(u_0)=f_0e^{u_0/p}$ and $\lambda(u_0)=\lambda_0e^{(1+s)u_0/p}$}

Here, $f_0$, $\lambda_0$, $p$ and $s$ are constants. In this case, we have the additional symmetry generator
\begin{equation}
X_6=x\partial_x+2p\partial_{u_0}+v_0\partial_{v_0}+2su_1\partial_{u_1}+(2s+1)v_1\partial_{v_1}
\label{gen4A}
\end{equation}
Performing a symmetry reduction corresponding to the subalgebra $\{X_6\}$, we obtain the solution
\begin{equation}
u_0=F(t)+2p\ln{x},\qquad v_0=xF_t,\qquad u_1=x^{2s}H(t),\qquad v_1=\dfrac{x^{2s+1}}{2s+1}H_t
\label{solution26}
\end{equation}
where
\begin{equation}
\int\dfrac{dF}{\sqrt{4p^2f_0e^{\frac{F}{p}}+K}}=t-t_0
\label{solution26A}
\end{equation}
and $H(t)$ satisfies the equation
\begin{equation}
H_{tt}+f_0e^{\frac{F}{p}}(-4s^2-6s-2)H-\dfrac{\lambda_0(1+s)}{p}e^{\frac{1+s}{p}F}\sqrt{4p^2+f_0e^{\frac{F}{p}}+K}(4ps+2p)=0
\label{solution26B}
\end{equation}
In the case where $\lambda_0=0$ and $K=0$, we obtain
\begin{equation}
F=-2p\ln{\left(\sqrt{f_0}(t_0-t)\right)},
\end{equation}
and equation (\ref{solution26B}) becomes
\begin{equation}\label{laterequation2B}
H_{tt}+f_0\left(-4s^2-6s-2\right)\left[-2p\ln{\left(\sqrt{f_0}(t_0-t)\right)}\right]^{-2}H=0.
\end{equation}
Therefore, we obtain the solution
\begin{equation}
u_0=-2p\ln{\left(\sqrt{f_0}(t_0-t)\right)}+2p\ln{x},\qquad v_0=\dfrac{2p}{t_0-t},\qquad u_1=x^{2s}H(t),\qquad v_1=\dfrac{x^{2s+1}}{2s+1}H_t
\end{equation}
where $H$ satisfies (\ref{laterequation2B}).

\subsection{The case where $f(u_0)=f_0(u_0+q)^{\frac{1}{p}}$ and $\lambda(u_0)=\lambda_0(u_0+q)^{\frac{1+s}{p}-1}$}

Here, $f_0$, $\lambda_0$, $p$, $q$ and $s$ are constants. In this case, we have the additional symmetry generator
\begin{equation}
X_6=x\partial_x+2p(u_0+q)\partial_{u_0}+(2p+1)v_0\partial_{v_0}+2su_1\partial_{u_1}+(2s+1)v_1\partial_{v_1}
\label{gen4B}
\end{equation}
Performing a symmetry reduction corresponding to the subalgebra $\{X_6\}$, we obtain the solution
\begin{equation}
u_0=x^{2p}F(t)-q,\qquad v_0=\dfrac{x^{2p+1}}{2p+1}F_t,\qquad u_1=x^{2s}H(t),\qquad v_1=\dfrac{x^{2s+1}}{2s+1}H_t
\label{solution27}
\end{equation}
where
\begin{equation}
\int\sqrt{\dfrac{2p+1}{2f_0p(4p^2-2p+1)F^{\frac{1}{p}+2}}}dF=t-t_0
\label{solution27A}
\end{equation}
and $H(t)$ satisfies the equation
\begin{equation}
\begin{split}
&H_{tt}-f_0\left(2s(2s-1)+2(2p-1)+8s+4p\left(\frac{1}{p}-1\right)\right)F^{\frac{1}{p}}H\\ &-\lambda_0\bigg{[}\left(\dfrac{1+s}{p}-1\right)\left(\dfrac{1+s}{p}-2\right)(4p^2)+\left(\dfrac{1+s}{p}-1\right)(2p)(2p-1)\\ &+2\left(\dfrac{1+s}{p}-1\right)(4p^2)+(2p)(2p-1)\bigg{]}F^{\frac{1+s}{p}-1}F_t=0
\end{split}
\label{solution27B}
\end{equation}
In the specific case where $p=\frac{1}{2}$, $s=-\frac{3}{2}$ and $t_0=0$, we obtain the explicit solution
\begin{equation}
\begin{split}
& u_0=x^{2p}\left(-\sqrt{\dfrac{2}{f_0}}\right)\left(\dfrac{1}{t}\right)-q,\qquad v_0=\dfrac{x^{2p+1}}{2p+1}\sqrt{\dfrac{2}{f_0}}\left(\dfrac{1}{t^2}\right),\\ & u_1=C_1t^{r_1}x^{2s}+C_2t^{r_2}x^{2s}+\dfrac{\sqrt{2}f_0^{\frac{3}{2}}\lambda_0x^{2s}t^2}{2(f_0-2)},\\ & v_1=\dfrac{x^{2s+1}}{2s+1}\left[C_1r_1t^{r_1-1}+C_2r_2t^{r_2-1}+\dfrac{\sqrt{2}f_0^{\frac{3}{2}}\lambda_0t}{2(f_0-2)}\right].
\end{split}
\label{solution27C}
\end{equation}
where $r_1=\frac{1}{2}\left(1+\sqrt{1+\frac{16}{f_0}}\right)$ and $r_2=\frac{1}{2}\left(1-\sqrt{1+\frac{16}{f_0}}\right)$.

\section{Concluding Remarks}

In this paper, the approximate symmetry analysis of a nonlinear wave equation with small dissipation has been performed. Based on the Lie symmetry approach, we determined subalgebras of dimension one and reduced the perturbed system of PDEs to systems of ODEs. These ODEs could often be explicitly integrated in terms of known functions or at least their singularity structure could be investigated using well-known methods. In particular, for ODEs of second and third order, it is possible to determine whether they are of the Painlev\'e type (i.e. whether all of their critical points are fixed and independent of the initial data). This approach has achieved a systematic classification of equations and invariant solutions from the group-theoretical point of view. Solutions obtained included elementary solutions (constant and algebraic solutions involving one or two simple poles), combinations of monomial powers of $x$ and $t$, solutions admitting damping and going to zero for large values of $t$ and solutions given by quadratures. This analysis can be applied to more general hydrodynamic systems admitting dissipation terms like viscosity and could lead to some new understanding of the problem of solving the Navier-Stokes system through the use of approximate symmetries.

\noindent {\bf Acknowledgements}\\
AMG's work was supported by a research grant from NSERC of Canada. AJH wishes to thank the Mathematical Physics Laboratory of the Centre de Recherches Math\'{e}matiques, Universit\'{e} de Montr\'{e}al, for the opportunity to participate in this research.

{}

\begin{appendix}
\section{Appendix: subalgebra classification for the integro-differential case}

The Lie symmetry subalgebra for the integro-differential case given in section 3 can be written as the semi-direct sum
\begin{equation}
{\mathcal L}=\{X_5,X_6\}\sdir \{X_1,X_2,X_3,X_4\}
\label{semidirectsum}
\end{equation}
The algebra $\{X_5,X_6\}$ is Abelian and its subalgebra classification is given by
\begin{equation}
\{0\},\qquad \{X_5\},\qquad \{X_6\},\qquad \{X_5+aX_6\}(a\neq 0),\qquad \{X_5,X_6\}
\label{abelianclassification}
\end{equation}
Using the method of splitting and non-splitting subalgebras as given in \cite{Winternitz}, we classify the one-dimensional subalgebras of the semi-direct sum (\ref{semidirectsum}). A basis element for each one-dimensional invariant subalgebra of ${\mathcal L}$ is transformed by the Baker-Campbell-Hausdorff formula in order to determine which other invariant subalgebras it is conjugate to. For instance, if we consider the subalgebra $X=\{X_1\}$ and take an arbitrary element of the group generated by ${\mathcal L}$, $e^Y$, where $Y$ is the generator
\begin{equation}
Y=\alpha X_1+\beta X_2+\gamma X_3+\delta X_4+\zeta X_5+\eta X_6
\label{groupelement}
\end{equation}
we obtain
\begin{equation}
e^YX_1e^{-Y}=X_1-\zeta X_1+\dfrac{\zeta^2}{2\!}-\ldots = e^{-\zeta}X_1
\label{BCHformula}
\end{equation}
so the subalgebra $\{X_1\}$ is conjugate only to itself. Applying this procedure to the other one-dimensional invariant subalgebras of ${\mathcal L}$, we obtain the following list of 63 one-dimensional subalgebras.

The following list constitutes the classification of the one-dimensional subalgebras of the symmetry Lie algebra for both cases of equation  (\ref{integralformA}) (where the symbol $X_6$ represents the symmetry generator (\ref{gen4A}) or the symmetry generator (\ref{gen4B}) respectively) into conjugacy classes.

\begin{displaymath}
\begin{split}
&{\mathcal L}_1=\{X_1\}, \hspace{5mm} {\mathcal L}_2=\{X_2\}, \hspace{5mm}
{\mathcal L}_3=\{X_1+\varepsilon X_2\}, \hspace{5mm}  {\mathcal L}_4=\{X_3\},  \hspace{5mm}
{\mathcal L}_5=\{X_3+\varepsilon X_1\}, \\ &  {\mathcal L}_6=\{X_3+\varepsilon X_2\},  \hspace{5mm}
{\mathcal L}_7=\{X_3+\varepsilon X_1+aX_2\}, \hspace{5mm} {\mathcal L}_{8}=\{X_4\},  \hspace{5mm}
{\mathcal L}_{9}=\{X_4+\varepsilon X_1\}, \\ &  {\mathcal L}_{10}=\{X_4+\varepsilon X_2\}, \hspace{5mm}
{\mathcal L}_{11}=\{X_4+\varepsilon X_1+aX_2\},  \hspace{5mm} {\mathcal L}_{12}=\{X_4+\varepsilon X_3\}, \\ & {\mathcal L}_{13}=\{X_4+\varepsilon X_3+aX_1\},  \hspace{5mm} {\mathcal L}_{14}=\{X_4+\varepsilon X_3+aX_2\}, \hspace{5mm}
{\mathcal L}_{15}=\{X_4+\varepsilon X_3+aX_1+bX_2\}, \\ & {\mathcal L}_{16}=\{X_5\},  \hspace{5mm} {\mathcal L}_{17}=\{X_5+\varepsilon X_1\},  \hspace{5mm}
{\mathcal L}_{18}=\{X_5+\varepsilon X_2\}, \hspace{5mm} {\mathcal L}_{19}=\{X_5+\varepsilon X_1+aX_2\}, \\ & {\mathcal L}_{20}=\{X_5+\varepsilon X_3\}, \hspace{5mm} {\mathcal L}_{21}=\{X_5+\varepsilon X_3+aX_1\}, \hspace{5mm}
{\mathcal L}_{22}=\{X_5+\varepsilon X_3+aX_2\}, \\ & {\mathcal L}_{23}=\{X_5+\varepsilon X_3+aX_1+bX_2\},  \hspace{5mm}
{\mathcal L}_{24}=\{X_5+\varepsilon X_4\}, \hspace{5mm} {\mathcal L}_{25}=\{X_5+\varepsilon X_4+aX_1\}, \\ & {\mathcal L}_{26}=\{X_5+\varepsilon X_4+aX_2\}, \hspace{5mm} {\mathcal L}_{27}=\{X_5+\varepsilon X_4+aX_1+bX_2\}, \hspace{5mm} {\mathcal L}_{28}=\{X_5+\varepsilon X_4+aX_3\}, \\ & {\mathcal L}_{29}=\{X_5+\varepsilon X_4+aX_3+bX_1\}, \hspace{5mm} {\mathcal L}_{30}=\{X_5+\varepsilon X_4+aX_3+bX_2\}, \\ &  {\mathcal L}_{31}=\{X_5+\varepsilon X_4+aX_3+bX_2+cX_1\}, \hspace{5mm} {\mathcal L}_{32}=\{X_6\}, \hspace{5mm} {\mathcal L}_{33}=\{X_6+\varepsilon X_1\}, \\ & {\mathcal L}_{34}=\{X_6+\varepsilon X_2\}, \hspace{5mm} {\mathcal L}_{35}=\{X_6+\varepsilon X_1+aX_2\}, \hspace{5mm} {\mathcal L}_{36}=\{X_6+\varepsilon X_3\}, \\ & {\mathcal L}_{37}=\{X_6+\varepsilon X_3+aX_1\},  \hspace{5mm} {\mathcal L}_{38}=\{X_6+\varepsilon X_3+aX_2\}, \hspace{5mm} 
{\mathcal L}_{39}=\{X_6+\varepsilon X_3+aX_1+bX_2\}, \\ & {\mathcal L}_{40}=\{X_6+\varepsilon X_4\}, \hspace{5mm}
{\mathcal L}_{41}=\{X_6+\varepsilon X_4+aX_1\}, \hspace{5mm}  {\mathcal L}_{42}=\{X_6+\varepsilon X_4+aX_2\}, \\ & 
{\mathcal L}_{43}=\{X_6+\varepsilon X_4+aX_1+bX_2\}, \hspace{5mm} {\mathcal L}_{44}=\{X_6+\varepsilon X_4+aX_3\},  \\ &
{\mathcal L}_{45}=\{X_6+\varepsilon X_4+aX_3+bX_1\}, \hspace{5mm} {\mathcal L}_{46}=\{X_6+\varepsilon X_4+aX_3+bX_2\},  \\ &
{\mathcal L}_{47}=\{X_6+\varepsilon X_4+aX_3+bX_1+cX_2\}, \hspace{5mm} {\mathcal L}_{48}=\{X_5+aX_6\}, \hspace{5mm} {\mathcal L}_{49}=\{X_5+aX_6+\varepsilon X_1\}, \\ &
{\mathcal L}_{50}=\{X_5+aX_6+\varepsilon X_2\},  \hspace{5mm} {\mathcal L}_{51}=\{X_5+aX_6+\varepsilon X_1+bX_2\}, \hspace{5mm} 
{\mathcal L}_{52}=\{X_5+aX_6+\varepsilon X_3\},  \\ & {\mathcal L}_{53}=\{X_5+aX_6+\varepsilon X_3+bX_1\}, \hspace{5mm}
{\mathcal L}_{54}=\{X_5+aX_6+\varepsilon X_3+bX_2\}, \\ & {\mathcal L}_{55}=\{X_5+aX_6+\varepsilon X_3+bX_1+cX_2\},  \hspace{5mm}
{\mathcal L}_{56}=\{X_5+aX_6+\varepsilon X_4\}, \\ & {\mathcal L}_{57}=\{X_5+aX_6+\varepsilon X_4+bX_1\}, \hspace{5mm} 
{\mathcal L}_{58}=\{X_5+aX_6+\varepsilon X_4+bX_2\}, \\ & {\mathcal L}_{59}=\{X_5+aX_6+\varepsilon X_4+bX_1+cX_2\}, \hspace{5mm}
{\mathcal L}_{60}=\{X_5+aX_6+\varepsilon X_4+bX_3\}, \\ & {\mathcal L}_{61}=\{X_5+aX_6+\varepsilon X_4+bX_3+cX_1\},  \hspace{5mm}
{\mathcal L}_{62}=\{X_5+aX_6+\varepsilon X_4+bX_3+cX_2\}, \\ & {\mathcal L}_{63}=\{X_5+aX_6+\varepsilon X_4+bX_3+cX_1+dX_2\},  \hspace{5mm}
\end{split}
\end{displaymath}
The subalgebra structure of the integro-differential case is far more extensive than that of the three cases analyzed in Section 2.

\end{appendix}

\end{document}